\documentclass[10pt,conference]{IEEEtran}
\usepackage{fancyhdr} 
\usepackage{cite}
\usepackage{amsmath,amssymb,amsfonts}
\usepackage{algorithm}
\usepackage[noend]{algcompatible}
\usepackage{graphicx}
\usepackage{textcomp}
\usepackage{color}
\usepackage{hyperref}
\usepackage{xspace}

\renewcommand{\COMMENT}[2][.5\linewidth]{%
  \leavevmode\hfill\makebox[#1][l]{//~#2}}

\newcommand{\registered}{\textsuperscript{\textregistered}\xspace}
\newcommand{\trademark}{\texttrademark\xspace}
\newcommand{\vtuneTM}{Intel\registered VTune\trademark Amplifier 2018\xspace}

\newcommand{\xeonReg}[0]{Intel\registered Xeon\registered}

\def\BibTeX{{\rm B\kern-.05em{\sc i\kern-.025em b}\kern-.08em
    T\kern-.1667em\lower.7ex\hbox{E}\kern-.125emX}}
    
\begin{document}

\title{ CosmoFlow: Using Deep Learning to Learn the Universe at Scale}

 \author{\IEEEauthorblockN{Amrita Mathuriya\IEEEauthorrefmark{1}, 
 Deborah Bard\IEEEauthorrefmark{2}, 
 Peter Mendygral\IEEEauthorrefmark{3}, Lawrence Meadows\IEEEauthorrefmark{1}, 
 James Arnemann\IEEEauthorrefmark{4}, \\ 
 Lei Shao\IEEEauthorrefmark{5}, 
 Siyu He\IEEEauthorrefmark{7}\IEEEauthorrefmark{2}\IEEEauthorrefmark{6},
 Tuomas K\"{a}rn\"{a}\IEEEauthorrefmark{1}, 
 Diana Moise\IEEEauthorrefmark{3}, 
 Simon J. Pennycook\IEEEauthorrefmark{5}, \\ Kristyn Maschhoff\IEEEauthorrefmark{3}, 
 Jason Sewall\IEEEauthorrefmark{5}, 
 Nalini Kumar\IEEEauthorrefmark{5}, 
 Shirley Ho\IEEEauthorrefmark{7}\IEEEauthorrefmark{2}\IEEEauthorrefmark{6}, 
 Michael F. Ringenburg\IEEEauthorrefmark{3}, Prabhat\IEEEauthorrefmark{2} 
 and Victor Lee\IEEEauthorrefmark{5}\\}
 
\IEEEauthorblockA{\IEEEauthorrefmark{1}Intel Corporation, 
 2111 NE 25th Ave, JF5, 
 Hillsboro, OR 97124, USA. Email: 
amrita.mathuriya@intel.com}
\IEEEauthorblockA{\IEEEauthorrefmark{2}Lawrence Berkeley National Laboratory, 
 1 Cyclotron Road, M/S 59R4010A, 
 Berkeley, CA 94720, USA}
\IEEEauthorblockA{\IEEEauthorrefmark{3}Cray Inc., 
 901 Fifth Avenue, Suite 1000, 
 Seattle, WA 98164, USA}
\IEEEauthorblockA{\IEEEauthorrefmark{4} U.C. Berkeley,
 Berkeley, CCA 94720, USA }
\IEEEauthorblockA{\IEEEauthorrefmark{5}Intel Corporation, 
 2200 Mission College Blvd., 
 Santa Clara, CA 95054, USA}
\IEEEauthorblockA{\IEEEauthorrefmark{6}McWilliams Center for Cosmology, 
Carnegie Mellon University, 5000 Forbes Avenue, 
 Pittsburgh, PA 15213, USA}
\IEEEauthorblockA{\IEEEauthorrefmark{7}Center for Computational Astrophysics, 
Flatiron Institute, 162 5th Ave, 
 New York, NY 10010, USA}
 }
\maketitle
\thispagestyle{fancy}
\lhead{}
\rhead{}
\chead{}
\lfoot{\footnotesize{
SC18, November 11-16, 2018, Dallas, Texas, USA
\newline 978-1-5386-8384-2/18/\$31.00 \copyright 2018 IEEE}}
\rfoot{}
\cfoot{}
\renewcommand{\headrulewidth}{0pt}
\renewcommand{\footrulewidth}{0pt}

\begin{abstract}
Deep learning is a promising tool to determine the physical model that describes our universe. 
To handle the considerable computational cost of this problem, we present CosmoFlow: a highly scalable deep learning application built on top of the TensorFlow framework.
CosmoFlow uses efficient implementations of 3D convolution and pooling primitives, together with improvements in threading for many element-wise operations, to improve training performance on Intel\registered Xeon Phi\trademark processors.  
We also utilize the Cray PE Machine Learning Plugin for efficient scaling to multiple nodes. 

We demonstrate fully synchronous data-parallel training on 8192 nodes of Cori with 77\% parallel efficiency, achieving 3.5 Pflop/s sustained performance. 
To our knowledge, this is the first large-scale science application of the TensorFlow framework at supercomputer scale with fully-synchronous training.
These enhancements enable us to process large 3D dark matter distribution and predict the cosmological parameters $\Omega_M$, $\sigma_8$ and $n_s$ with unprecedented accuracy.

\end{abstract}

\begin{IEEEkeywords}
Cosmology, Deep Learning, Machine Learning, TensorFlow, High Performance Computing 
\end{IEEEkeywords}

\section{Overview}
\label{sec:overview}

\subsection{Deep Learning for Science}

Deep Learning is a powerful technique for learning the relationships between variables in complex data. 
It has been widely developed and adopted in commercial applications to solve classification and regression problems. 
Many science areas face similar classes of problems, with complex data that contains features that cannot be readily extracted through traditional statistical methods. 
Deep learning techniques are poised to have a major impact on many scientific domains~\cite{oreilly}.
However, deep learning with scientific data has unique challenges that are typically not faced in the commercial world. 
Scientific data is often complex (multi-dimensional (3D, 4D) with several channels) and voluminous (TB-PBs in size). 
To be relevant for scientific problems and allow fast turnaround of the exploration of ideas, deep learning frameworks need to efficiently process multi-dimensional data at scale.

\subsection{Deep Learning for Cosmology}
The nature of dark energy is one of the most exciting and fundamental questions facing scientists today. 
Dark energy is the unknown force that is driving the accelerated expansion of the universe, and is the subject of several current and future experiments that will survey the sky in multiple wavelengths (for example LSST\footnote{https://www.lsst.org/}, DESI\footnote{http://desi.lbl.gov/}, DES\footnote{https://www.darkenergysurvey.org/}, WFIRST\footnote{https://wfirst.gsfc.nasa.gov/}). 
We cannot measure dark energy directly - we can only observe the effect it has on the observable universe. 
The interplay of gravity (pulling matter together) and dark energy (expanding space itself) is encoded in the distribution of matter in the universe. 
Cosmologists typically characterize this distribution using statistical measures of the structure of matter - measures of its ``clumpiness'' - in the form of two- or three-point correlation functions \cite{Dodelson2003} or other reduced statistics.
Methods that capture all features in the distribution of matter (such as deep learning networks)  could give greater insight into the nature of dark energy.

\begin{figure}
\centering
\includegraphics[width=\columnwidth]{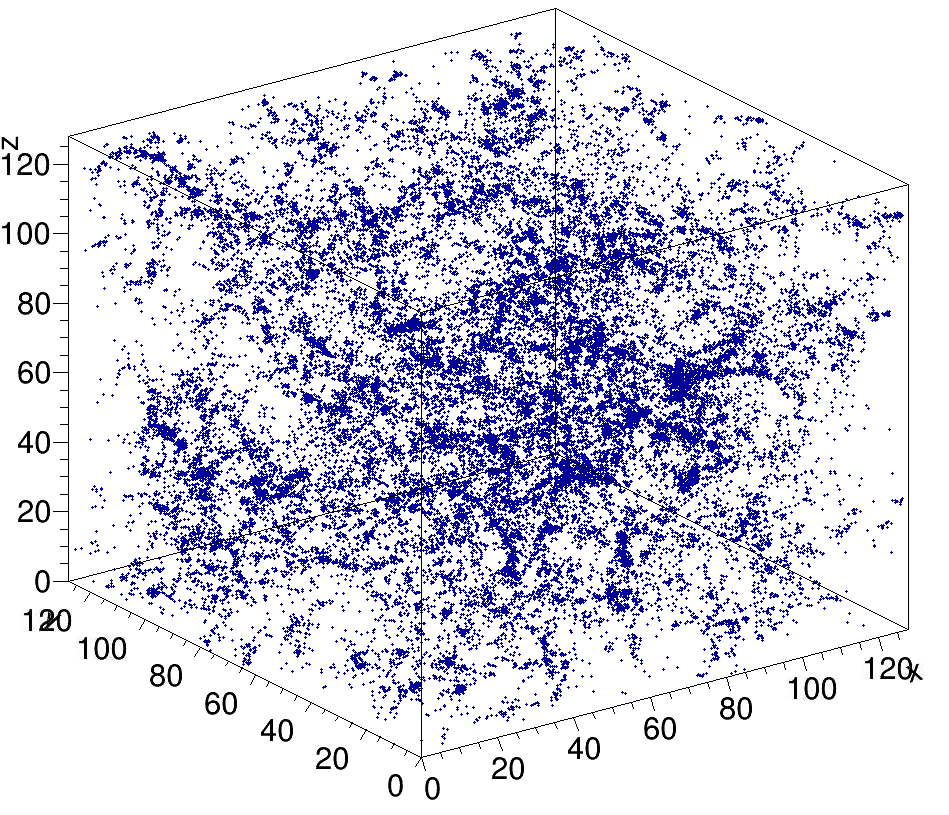}
    \caption{Example simulation 256$Mpc/h ^3$, 128$^3$ voxel sub-volume, used as input to the CosmoFlow network. This sub-volume is taken from the full 512$Mpc/h ^3$ simulation of dark matter in the universe, evolved over 3 billion years to a redshift of 0 (i.e. today). }
\label{fig:sim}
\end{figure}

\subsection{Contributions}
The CosmoFlow project aims to process large 3D cosmology datasets on modern HPC platforms. Our specific contributions are as follows:

\begin{itemize}
\item We adapt the deep learning network described by Ravanbakhsh et al. (2017)~\cite{siamak2017} to a scalable architecture for a larger problem size of $128^3$ voxels and predict three cosmological parameters. 
We perform simulations and generate the cosmology dataset used in this work with the 3 parameter variations.
\item We implement efficient primitives in MKL-DNN\cite{mkldnnSite} for 3D convolutional neural networks, which are used in an optimized TensorFlow~\cite{abadi2016tensorflow} framework for CPU architectures.
\item We utilize the Cray Programming Environments Machine Learning Plugin (CPE ML Plugin) to efficiently scale deep learning training on supercomputers via MPI.
\item We demonstrate for the first time fully synchronous data parallel training with high efficiency at supercomputer scale and achieve sustained 3.5 Pflop/s single precision performance on 8192 Cori Intel\registered Xeon Phi\trademark processor (KNL) nodes. We discuss the training methodology at 2048 and 8192 nodes for convergence and trained model predictions. The single node and scaling performance improvements enable us to process a large 3D dark matter distribution and predict the cosmological parameters $\Omega_M$, $\sigma_8$ and $n_s$ with unprecedented accuracy.
\end{itemize}

\section{State of the Art}

\subsection{Deep Learning in Cosmology}

In recent years deep learning methods have started to be used in astronomy, including for the prediction of galaxy morphology~\cite{Dieleman2015} and generating models for galaxies images~\cite{Siamak2016}. 
Today, deep learning is starting to gain traction in cosmology, as the ability of these algorithms to learn features in the universe becomes apparent. 
For example, deep learning is being used in cosmology simulations to reduce the computational load of expensive simulations of the universe. 
Deep Learning was used by Lucie-Smith et al. (2018)~\cite{LucieSmith2018} to predict whether regions of an initial density field of dark matter will collapse into dark matter halos and therefore form galaxies. 
Generative Adversarial Networks were used by  Mustafa et al. (2017)~\cite{Mustafa2017} to generate 2D mass maps of the sky (via gravitational lensing~\cite{Dodelson2003}) for specific cosmological models, indistinguishable by all standard statistical metrics from maps produced by full-scale N-body simulations. 

We base this paper on the work presented by Ravanbakhsh et al. (2018)~\cite{siamak2017}, which was the first attempt to learn cosmological parameters directly from the distribution of matter. 
The authors address a regression problem, where the best-fit cosmological parameters are estimated given an observed three-dimensional dark matter distribution. 
This work was the first to demonstrate that deep learning can be used to significantly improve cosmological parameter estimation, seeing a reduction in relative error of up to a factor of 3 compared to traditional statistical metrics.

Others have used deep learning to estimate the parameters which describe the physical model of the universe from 2D data. 
Schmelzle et al. (2017)~\cite{Schmelzle2017} used deep learning with data from 2D mass maps to address a classification problem in estimating two cosmological parameters, for 5 specific cosmological models. 
The authors found that the deep convolutional neural network (CNN) they developed did significantly better at estimating the cosmological model behind the mass maps than traditional  metrics, indicating that their network identified useful features in these maps beyond those traditionally defined by scientists. 
They also addressed the issue of training networks to deal with noise in real scientific data, an aspect which we expect will accrue more attention as the use of advanced machine learning techniques is increasingly applied to noisy, complex scientific data. 

We also note that the Schmelzle et al. (2017)~\cite{Schmelzle2017} required 500 wall-clock hours to train their network on 12,500 2D maps, using a single NVIDIA P100 GPU. 
This is a common hindrance in the use of deep learning methods for scientific analysis; a turnaround time of 20 days does not enable the rapid development and testing of research ideas. 
HPC can be used to address this issue and speed time to discovery.

\subsection{Deep Learning on single Node}

In deep neural networks (DNN), the most compute intensive calculations occur in the convolution kernels, which are similar in nature to matrix-matrix multiplication (and in some prior works, SGEMM has been used for computing convolutions~\cite{chetlur2014cudnn}).
Device manufacturers have been augmenting programmable processors (such as CPU and GPU) with accelerators to speed up convolution operations~\cite{jouppi2017datacenter, nervanaChip, NVdiaTensorRT}.
These accelerators are capable of providing $> 10^{12}$ AIOps/sec compute with specialized libraries such as MKL-DNN~\cite{mkldnnSite} or CuDNN~\cite{cudnnSite}.   
However, training a complex DNN requires up to an Exaflop of compute~\cite{resnet} and these accelerators do not have the energy efficiency to solve this training problem at large scale (with limited power budget).  

TensorFlow~\cite{abadi2016tensorflow} is an open source software library for numerical computation using data-flow graphs, operating on multidimensional data arrays referred to as `tensors'. It is commonly used for machine learning applications, and is widely popular among the scientific community.
We focus our work on accelerating 3D CNNs for CPU architecture in TensorFlow framework, optimizing not just the convolutional kernels but all stages involved in the full network topology.  

\subsection{Deep Learning on multiple nodes}
\label{sec:multi}

Deep learning is inherently computationally demanding.
HPC systems and methods are being adopted in two ways to reduce time-to-solution and to scale to larger problems sizes.

In one approach, HPC is used to enable a massively-parallel hyperparameter search. 
For example, Young et al. (2017)~\cite{Young2017} scanned the possible space of network hyperparameters, using up to 18,000 GPUs. 
In this approach, each node in the HPC system independently trains a different network, and aggregates the results to determine which network design in the ensemble gives the best results.

In the second approach, an HPC system can be used to train a single network. In a model-parallelism method different nodes optimize different parts of the same network using the same data.
This method is important if the model size is too large or too complex to fit on a single compute node. 
In a data-parallel training method (which we utilize in this paper), different nodes train the same model on different subsets of the data. After every training step the nodes pool their results to decide what should happen in the subsequent step. 
The global batch size is, therefore, the sum of batch sizes over all compute nodes.
This data-parallel method is particularly suitable for datasets that are either too large to fit in the memory of a single compute node or will take too long to train using a single node, for example in Kurth et al. (2017)~\cite{kurth2017deep}. 

Synchronous approaches for multi-node deep learning use collective global reductions to force the ensemble of nodes to perform every parameter update at the same time. 
Attempts to scale synchronous methods have found that the scalability slows after a few hundred nodes~\cite{Iandola2015,Das2016,Pan2017,Zheng2016}.
This is due to a number of factors, including the observation that large batch sizes can slow down convergence without careful tuning of the optimizer, or that a single slow node can significantly reduce the aggregate performance.
In this work, we further explore the impact of IO variability on this ``straggler" effect.

In an asynchronous approach, by contrast, each node independently contributes to a central parameter server~\cite{kurth2017deep, Tsitsiklis1986,Niu2011,Dean2012}. 
Such systems need more iterations-to-solution, since each node is always working on a slightly out-of-date version of the best parameter estimates, resulting in training inefficiencies.

TensorFlow by default uses the GRPC\footnote{https://grpc.io/docs/} protocol for communication.
This implements a centralized master-slave-based algorithm for an AllReduce operation of gradients. Mathuriya et al. (2017)~\cite{mathuriya2017scaling} showed that this approach and implementation does not scale to large node counts due to algorithmic inefficiencies and socket-based communication. The CPE ML Plugin, used in this work, uses MPI-based scalable algorithms, specifically designed and optimized for DL workloads, to overcome this issue.  An alternative parallelization framework is Horovod \cite{horovod}.  It uses general purpose MPI collectives for gradient aggregation.  Horovod is an option for scientists looking for portability to any system that supports MPI.

To the best of our knowledge, this work is the first large scale science application of the TensorFlow framework to efficiently run at supercomputer scale with fully-synchronous training and is also the first to process 3D spatial data volumes at this scale.

\section{Innovations}
\label{sec:innovations}
In this work we present CosmoFlow, a highly optimized and scalable deep learning application that predicts the cosmological parameters that describe the nature of the universe given the 3D matter distribution.
Our deep learning stack uses Python as the front-end and is built on top of the TensorFlow framework. 
We optimize 3D CNN performance in MKL-DNN~\cite{mkldnnSite} and TensorFlow for KNL processors.
In this section, we describe the methods employed to accomplish our performance goals.

\subsection{3D convolutional network and dataset}
Our work is based upon the pioneering work done by Ravanbakhsh et al. (2017)~\cite{siamak2017}, where the authors demonstrated that a convolutional neural network can learn cosmological parameters using 3D data. 
That work used $64^3$-voxel simulation volumes as training data and predicted two parameters. 
Scaling up the network presented a computational bottleneck. 
The aim of this paper is to predict three cosmological parameters by adapting the network topology developed by
Ravanbakhsh et al. (2017)~\cite{siamak2017} 
and scale the problem to a dataset an order of magnitude larger in physical volume (i.e. 128$^3$ voxels), and an additional order of magnitude increase in training samples. 
Details of the datasets we generated and the cosmology parameters are described in section~\ref{sec:data}.

\begin{figure*}
\centering
\includegraphics[width=2\columnwidth]{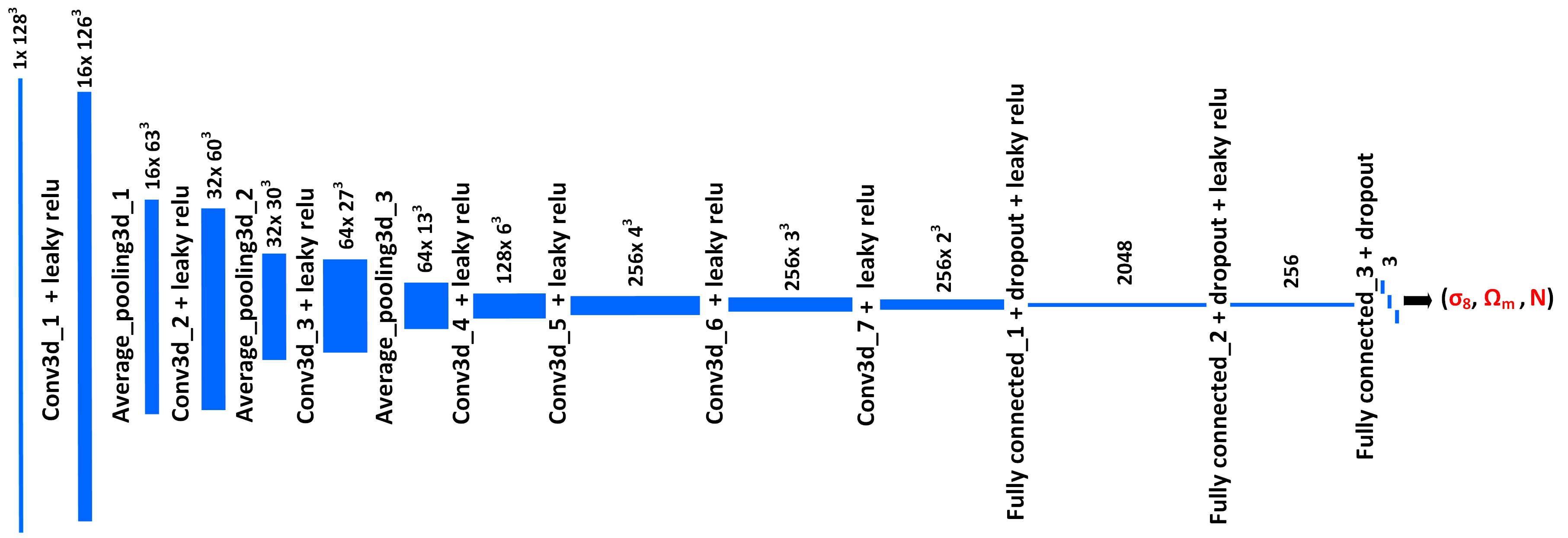}
\caption{CosmoFlow network topology, showing the different network layers and data sizes at each layer.}
\label{fig:network topology}
\end{figure*}

CosmoFlow's network topology consists of 7 convolution layers and 3 fully-connected (FC) layers, as shown in Figure~\ref{fig:network topology}. 
Similar to the topology presented by Ravanbakhsh et al. (2017)~\cite{siamak2017}, our network contains convolution layers followed by average pooling layers with stride of (2,2,2) to reduce the spatial dimensions of the inputs, while doubling the number of output channels. 
Increasing the spatial dimension of inputs directly impacts the number of input neurons in the first FC layer, which in turn can increase the number of parameters in the network significantly. 
To cope with the problem size going from $64^3$ to $128^3$ pixels, we include an additional convolution layer, doubling the number of output channels, and an average pooling layer which down-samples the inputs with the stride of (2,2,2). 
This allows us to keep the overall topology the same while keeping the number of network parameters manageable with the increased problem size. We also update the number of output neurons in the last FC layer to 3 corresponding to the three cosmological parameters predicted. All convolution and FC layers use leaky Relu as their activation function.

In addition, we optimize CosmoFlow's topology in two ways for performance considerations.
\begin{itemize}
\item We increase the number of output channels 
for all convolution layers to be a multiple of 16 
to allow for efficient vectorization over the channel dimension.
\item We remove batch-norm layers from the topology for efficient scaling and compute performance. We use a batch size of one for all our experiments, and do not see accuracy degradation with batch-norm removal. 
\end{itemize}

\subsection{SGD Optimizer}
We use fully synchronous training for each of the runs. All runs use a batch size of 1 for each MPI rank, which gives an effective batch equal to the number of MPI ranks with 1 MPI rank per node.  For all training runs, we use Adam~\cite{Adam} as our base optimizer with $\beta_{1} = 0.9$, $\beta_{2} = 0.999$, $\hat{\epsilon} = 10^{-8}$.
We combine Adam with the Layer-wise Adaptive Rate Control (LARC)~\cite{larc-paper} technique and a polynomial (power=1) learning rate decay schedule.  LARC is a variant of Layer-wise Adaptive Rate Scaling (LARS)~\cite{lars-paper}. LARS/LARC adjust the magnitude of the update with respect to the weight norm for each layer for better control of training speed and stability.  On top of LARS, LARC includes a clip operation for each layer such that the effective learning rate will not exceed the nominal learning rate for Adam.

The polynomial learning rate decay enables larger learning rates early in training, and thus faster training, but slows training down to aid in convergence to a local minima to overcome the training difficulty typically observed at large effective batch sizes.  Parameters $\vec{v}_{l,t}$ in a layer $l$ at step $t$ with gradients $\vec{g}_{l,t}$ are updated according to
\begin{align*}
\eta_{t} = (\eta_{0} - \eta_{min}) \times (1 - \frac{t}{t_{decay}}) + \eta_{min} \\
{v}_{l,t} = ||\vec{v}_{l,t}||_{2}, {g}_{l,t} = ||\vec{g}_{l,t}||_{2} \\
\eta_{l,t}^{*} = 
\begin{cases}
\begin{split}
0.002 \frac{{v}_{l,t}}{{g}_{l,t}}, & \text{  for } {v}_{l,t} \neq 0 \text{ and } {g}_{l,t} \neq 0 \\
   6.25\times 10^{-5} & \text{  otherwise}
\end{split}
\end{cases} \\
\eta_{l,t}^{\dagger} = \text{min}(\eta_{l,t}^{*}, 1) \\
\vec{g}^{*}_{l,t} = \eta_{l,t}^{\dagger} \vec{g}_{l,t} \\
\vec{v}_{l,t+1} = \text{Adam}(\vec{v}_{l,t}, \vec{g}^{*}_{l,t}, \eta_{t}),
\end{align*}
where $\eta_{t}$ and $\eta_{l,t}^{\dagger}$ is global learning rate at step $t$ and the clipped local learning rate for layer $l$ at step $t$, respectively and
$\eta_{0} = 2\times10^{-3}$, $\eta_{min} = 10^{-4}$. 

\subsection{Single-node optimizations}

To improve single-node performance, we analyze the CosmoFlow topology with the \vtuneTM~\cite{vtuneCite} update 2 profiler. 
We identify the primary computational hotspots, which include 3D convolution and pooling operators. We optimize these operators within the MKL-DNN library. Subsequently identified hotspots include element-wise operations in TensorFlow, for which we implement simple loop-level parallelism with OpenMP.
This makes 
the threading scheme used in TensorFlow more compatible with MKL-DNN's OpenMP threading.

\begin{algorithm}
\caption{Pseudocode for 3D forward convolution $DST = \text{conv}(SRC, W)$.
The input ($SRC$) and output ($DST$) arrays have been blocked by 16 channels.
The weight array ($W$) has been blocked by 16 input and output channels.
Variables $s,d,w$ are pointers to an element in the larger arrays.
For the sake of simplicity we assume that stride is one and output width is a multiple of 28 voxels. \label{alg:fwd_conv3d}}
 \begin{algorithmic}[1]
 \small
    \REQUIRE $SRC \in \mathbb{R}^{ICB \times ID \times IH \times IW \times 16}$ 
    \REQUIRE $DST \in \mathbb{R}^{OCB \times OD \times OH \times OW \times 16}$ 
    \REQUIRE $W \in \mathbb{R}^{OCB \times ICB \times KD \times KH \times KW \times 16 \times 16}$ 
    \STATE \textbf{for} $ocb = 1\cdots OCB$ \textbf{do} \COMMENT{Output channel block}
    \STATE \textbf{for} $icb = 1\cdots ICB$ \textbf{do} \COMMENT{Input channel block}
    \STATE \textbf{for} $od = 1\cdots OD$ \textbf{do} \COMMENT{Output depth}
    \STATE \textbf{for} $oh = 1\cdots OH$ \textbf{do} \COMMENT{Output height}
    \FOR{$owb=1\cdots OWB$} \COMMENT{Output width block}
    
      \STATE $d \leftarrow DST[ocb, od, oh, 28owb, 0]$
      \STATE \textbf{for} $kd = 1\cdots KD$ \textbf{do} \COMMENT{Kernel depth}
      \STATE \textbf{for} $kh = 1\cdots KH$ \textbf{do} \COMMENT{Kernel height}
      \FOR{$kw = 1\cdots KW $} \COMMENT{Kernel width}

      \STATE $s \leftarrow SRC[icb, od + kd, oh + kh, 28owb + kw, 0]$
      \STATE $w \leftarrow W[ocb, icb, kd, kh, kw, 0, 0]$
      \STATE \textbf{for} $ow=1\cdots 28$ \textbf{do} \COMMENT{Output width}
      \STATE \textbf{for} $oc=1\cdots 16$ \textbf{do} \COMMENT{Output channel}
      \FOR{$ic=1\cdots 16$} \COMMENT{Input channel}
        \STATE $d[16ow + oc] \leftarrow w[16ic + oc] s[16ow + ic]$
      \ENDFOR
      \ENDFOR
    \ENDFOR
\end{algorithmic} 
\end{algorithm}
We use a direct convolution algorithm, where the computation of each output element involves a 4-dimensional inner-product over the weight array and input channels.
We optimize 3D convolution operators by vectorizing the innermost loops, applying cache- and register blocking, and threading. 
Algorithm \ref{alg:fwd_conv3d} outlines the forward convolution procedure for layers where the number of input and output channels is a multiple of 16. The input, output, and weight arrays are blocked by channels. 
Block size is set to 16 to match the target machine's single-precision SIMD width. 
Additionally, we block the output width dimension by 28 voxels. This results in three innermost loops with sizes 28, 16, and 16. 
These loops are completely unrolled and vectorized with AVX512 SIMD instructions, making use of all the 32 SIMD registers available. We employ a just-in-time-assembly framework \cite{xbyak} to generate the desired AVX512 instructions for the inner loops. Thread decomposition is done over the output voxel space, each thread writing to a separate block.

The backward propagation pass involves two operators: the backward data operator that propagates the difference signal to the input layer, and backward weight operator that computes the weight difference signal. These operators are optimized with a similar strategy by blocking the channels and using SIMD vectorization.

The backward weights operator is equivalent to a forward convolution with large inputs and kernels and produces a small output tensor. 
When the number of channels is small, threading is done over the output image voxels: each thread operates on a distinct region of the output image. 
Threads accumulate weights in thread-private scratch arrays before combining them all at the end via a parallel reduction. 
On layers with sufficiently many input/output channels, we divide work amongst threads to minimize the reduction overhead: channel blocks are first assigned to teams of threads, then voxels are assigned to threads in the team. In cases where there are a sufficient number of channel blocks (e.g. the last few layers of the CosmoFlow topology), we do not parallelize across voxels at all and are able to skip the reduction entirely.

Average pooling is a special case of the convolution operator: each channel is averaged separately, and the weights array is a constant (each element being $1/(KS)^{3}$ for a kernel of size $KS$). The lower arithmetic intensity results in the pooling operator being bandwidth-bound. Our optimizations include the aforementioned 16 channel blocking, AVX512 vectorization, and just-in-time-assembly. Threading is done over the output voxel space.

\subsection{Multi-node scaling}

The CPE ML Plugin\footnote{The CPE ML Plugin is included with the Cray Urika{\textregistered}-XC package.} (available for Cray systems) is an MPI-based, framework-independent plugin for parallelizing the training of deep learning networks.  
It includes easy-to-use C/Python interfaces and provides highly optimized communication primitives designed for deep learning workloads. 
The API and communication algorithms in the CPE ML Plugin are specifically designed and optimized for deep learning training.

Data-parallel training is an ideal approach for this work due to the volume of training data.  Our approach is fully synchronous training (hereafter Synchronous Stochastic Gradient Descent or SSGD) as given in Algorithm \ref{alg:parallel_training}.

\begin{algorithm}
\caption{Pseudocode for data-parallel synchronous training algorithm. The CPE ML Plugin is represented by $mc$, and the gradient aggregation function is $mc.gradients()$. \label{alg:parallel_training}}
 \begin{algorithmic}[1]
 \small
      \REQUIRE N = total number of epochs
      \REQUIRE n = total number of training samples
      \REQUIRE k = number of MPI ranks
      \FOR{$\text{epoch}=1\cdots N$}
      \FOR{$\text{step}=1\cdots n/k$}
      \State $g_{step} \gets $compute\_gradients(local\_batch$_{step}$)
      \State $G_{step} \gets $mc.gradients($g_{step}$)
      \State $loss_{step} \gets $apply\_gradients($G_{step}$)
      \ENDFOR
      \ENDFOR

\end{algorithmic}
\end{algorithm}

The CPE ML Plugin reduces the ``straggler'' effect in SSGD by using non-blocking MPI communication to hide timing imbalances across processes through the stages of the reduction.  There are no unique processes (e.g. parameter servers, backup workers) in its parallel design.  Every MPI rank is a worker computing gradients.  This eliminates redundant or wasted resources, simplifies its usage and requires little to no extra tuning by the user.

The CPE ML Plugin uses a pool of helper threads for communication.  Threads can be organized into teams where each team progresses a gradient aggregation independently.  The number of teams and threads per team is tuned by the user when initializing the CPE ML Plugin.  Each thread in a team progresses a portion of gradient aggregation independently with infrequent synchronization across threads.  Using multiple threads for communication can increase network utilization, in particular on Intel\registered Xeon Phi\trademark processor architectures \cite{cori_network, wombat_knl}.  Four helper threads in a single team are used for the runs on Cori and two helper threads in a single team are used for Piz Daint runs in this work.

For this work no modifications are made to the TensorFlow source to make use of the CPE ML Plugin.  Instead we include calls to it in the Python training script. The CPE ML Plugin adds itself to the TensorFlow graph with a custom TensorFlow operation\footnote{Documentation on custom TensorFlow operations is found at \url{https://www.tensorflow.org/extend/adding_an_op}.}.  This allows direct access to TensorFlow memory and minimizes unnecessary copies of gradient data.

\section{Systems and Datasets}
\label{sec:}
In this section we describe the dataset configurations that were used to perform the scaling measurements. We also describe the two HPC systems -- Cori and Piz Daint -- on which these computations were performed.

\subsection{Description of the Cori System}
\label{sec:cori}
The CosmoFlow code is run on the Cori system at the National Energy Research Scientific Computing Center (NERSC) at Lawrence Berkeley National Laboratory. 
Cori is a Cray XC40 system featuring 2,004 nodes of \xeonReg
Processor E5-2698 v3 (``Haswell'') and 9,688 nodes of Intel\registered Xeon Phi\trademark Processor 7250 (KNL). 
All computations presented here are performed on KNL nodes. 
Each of these nodes contains 68 cores (each supporting 4 simultaneous hardware threads), 16 GB of on-package, multi-channel DRAM (``MCDRAM''), and 96 GB of DDR4-2400 DRAM. 
Cores are connected in a 2D mesh network with 2 cores per tile, and 1 MB cache-coherent L2 cache per tile.
All measurements reported in this work are performed
with the MCDRAM in ``cache'' mode (configured as a transparent, direct-mapped cache to the DRAM).
Each core has 32 KB instruction and 32 KB data in L1 cache. 
The nodes are connected via the Cray Aries interconnect.

In addition, the Cori system contains 288 nodes as part of the Cray DataWarp system (also known as the ``Burst Buffer"). 
Each DataWarp node contains 2$\times$3.2TB SSDs, giving a total of 6.4TB per node and a system total of roughly 1.8PB of SSD storage. 
The Burst Buffer is measured to give up to 1.7TB/sec read/write performance and over 28M IOP/s, making it one of the fastest IO systems in the world. 
We stripe our data over 125 DataWarp nodes with the default stripe size of 8MB. 

Cori also has a Sonnexion 2000 Lustre filesystem, which consists of 248 Object Storage Targets (OSTs) and 10,168 disks, giving nearly 30PB of storage and a maximum of 700GB/sec IO performance. 
We stripe our data over 64 OSTs on Cori Lustre with the default stripe size of 1 MB.

\subsection{Description of the Piz Daint system}\label{sec:piz-daint}
Piz Daint is a Cray XC50 system located at the Swiss National Supercomputing Center (CSCS) with 1,431 dual socket nodes populated with \xeonReg E5-2695 v4 processors, and 5,320 hybrid nodes each with one \xeonReg E5-2690 v3 processor and one NVIDIA P100 (PCIe) GPU.  It has a theoretical peak of more than 25 Pflop/s.  The work presented here uses the hybrid nodes, which each have 64 GB of DRAM and 16 GB of high bandwidth memory (HBM) on the GPU.

The Piz Daint Sonexion 3000 Lustre filesystem consists of 40 OSTs, has a capacity of 6.2 PB and an aggregate peak bandwidth of 112 GB/s.  In all of our tests we striped training data over 16 OSTs.  Piz Daint also has a DataWarp  filesystem, but we do not make use of it for this work due to availability issues.

\subsection{Simulation Data}
\label{sec:data}
We train the CosmoFlow network using dark matter N-body simulations produced using the {\it MUSIC} and {\it pycola} packages. 
{\it MUSIC}\footnote{https://www-n.oca.eu/ohahn/MUSIC/}~\cite{music2011} (MUlti-Scale-Initial-Conditions) is used to create the initial conditions for the simulations. 
A slightly modified version of {\it pycola}\footnote{https://bitbucket.org/tassev/pycola/}~\cite{cola2013,cola2015} is used to create the dark matter simulations. 
{\it pycola} is a multithreaded Python/Cython N-body code, implementing the Comoving Lagrangian Acceleration (COLA) method in the temporal and spatial domains. 
The COLA approach preserves N-body accuracy at large scales, but is significantly faster to run than a traditional N-body simulation code. 
This allows us to run a large suite of fast but accurate simulations to be used as training data. 

We use simulation volumes of 512h$^{-1}$Mpc$^3$ containing 512$^3$ dark matter particles, which are evolved from the random density fluctuations provided by the MUSIC initial conditions to a redshift ($z$) of zero (i.e. to the present day). We use the $z=0$ snapshot for our training dataset.
We run 12,632 simulation boxes, varying $\Omega_M$, $\sigma_8$ and $n_s$. 
$\Omega_M$ describes the proportion of matter in the universe. We assume a flat geometry for universe, i.e. the sum of the contributions of matter and dark energy to the energy density of the universe $\Omega_M + \Omega_{\Lambda} = 1$. 
$\sigma_8$ defines the amplitude of mass fluctuations in the universe at a distance scale of 8Mpc/h, and effectively sets the scale of the matter density distribution we see in the universe today. $n_{s}$ is the scalar spectral index of the spatial curvature of a comoving slicing of
the space-time. 
The best estimates of these parameters come from a combination of observations. 
We base the parameter ranges we use in our work on the measurements made by the Planck collaboration from the Cosmic Microwave Background~\cite{Planck2016}, which give $\Omega_M = 0.3089\pm 0.0062, \sigma_8 = 0.8159\pm 0.0086$, $n_s = 0.9667\pm 0.0040$. 
Accordingly, we use an evenly sampled set of random parameters in the ranges $(0.25<\Omega_M<0.35), (0.78<\sigma_8<0.95),$ {$(0.9<n_s<1.0)$} in our simulations.

Each of our 512h$^{-1}$Mpc$^3$ simulations cubes contains 512$^3$ particles. This volume is histogrammed into a 256$^3$-voxel 3D histogram of particle counts using the python function numpy.histogramdd, and then split into 8 sub-volumes. 
This gives us 8$\times$128$^3$-voxel sub-volumes per simulation (each 256h$^{-1}$Mpc$^3$ in spatial extent), for a total of 
101,056 subvolumes. 
An example of this volume is shown in Figure~\ref{fig:sim}. 
We choose this simulation size based on the scale of the physical structures in the simulations. 
Galaxy clusters, which are widely regarded as sensitive cosmological probes, are typically around 10h$^{-1}$Mpc$^3$ in size and separated by around 50h$^{-1}$Mpc$^3$~\cite{Mo}.  
Our data volumes of 256h$^{-1}$Mpc$^3$ (with a resolution of 2h$^{-1}$Mpc$^3$) represent a volume of space that allows multiple clusters to be present at a resolution that will include features such as dark matter filaments. 

We set aside 150 simulations (i.e. 1200 sub-volumes) to hold as the validation data, and 50 simulations (i.e. 400 sub-volumes) to hold as the test data. 
This leaves us with
99,456 sub-volumes for the training dataset 
which we duplicate once to augment our training dataset. 

The TFRecord file format is a simple record-oriented binary format commonly used in TensorFlow. 
We randomly assign the training sub-volumes to TFRecord files for training. 
We do not randomize or repeat the validation and test TFRecords. 
Each TFRecord contains 64 samples and is 512MB in size. The total amount of TFRecord data is 1.4TB.

\subsection{Libraries and Environment}
The TensorFlow code used in this work is an extension of the 1.5 release (r1.5) branch~\cite{tfCode}. 
On Cori, TensorFlow is compiled with the GNU compiler gcc 7.2.0, using bazel version 0.11.1 and MKL-DNN with the \texttt{-config=mkl} flag. Bazel copt flags are ``\texttt{mfma, mavx2, O3, ggdb3, march=broadwell} and \texttt{DINTEL\_MKL\_DNN}'' and we use the cxxopt option is \texttt{D\_GLIBCXX\_USE\_CXX11\_ABI=0}.
We run TensorFlow with \texttt{inter\_op} and \texttt{intra\_op} thread settings equal to one.
We use the \vtuneTM~\cite{vtuneCite} update 2 profiler to identify primary computational hotspots on Cori.
Our MKL-DNN development is based on the master branch (tag ~\cite{mkldnnCode}). 
On Cori, at run time, we use the Intel\registered distribution for Python version 2.7, unset \texttt{OMP\_NUM\_THREADS} variable, and set the thread affinity using ``\texttt{KMP\_AFFINITY=compact}'' and {``\texttt{KMP\_HW\_SUBSET=$64C@2,1T$}''. We use the CPE ML Plugin version 1.1.0 along with Cray MPICH version 7.6.2. We run the code with 1 MPI process per compute node.
We also run CosmoFlow on Piz Daint, where we use cuDNN v7~\cite{cudnnSite} and CUDA 8 with publicly available TensorFlow r1.5 version for NVIDIA P100 (PCIe) GPUs with 1 MPI rank per node.
We use four and two helper threads of CPE ML Plugin on Cori and Piz Daint respectively.

\section{Experiments and Results}
\label{results}
A key objective of this work is to achieve convergence at extreme scale with high compute and scaling efficiency. 
Convergence of the deep learning algorithm is directly affected by the global batch size,  which is the summation of mini-batch size processed by each node, and which increases proportionally with the node count. 
We therefore process a mini-batch size of one on a single node to keep the global batch size manageable at the full supercomputer scale.

In this section, we describe the methodology used for calculating the floating point operations rate (flop/s) and present results and analysis of single node and scaling runs.
We also describe our full-scale runs on Cori, which are targeted for high performance  with convergence.

\subsection{Workflow and Network's FLOP rate computation}
As stated in Section~\ref{sec:innovations}, the CosmoFlow network consists of 7 convolution and 3 fully-connected layers, together with three average pooling layers. 
The last four convolution layers have relatively little computation due to the smaller input sizes. The network consists of slightly more than seven million parameters. Both the input dataset and the weights use 32-bit single precision floating point format. With a mini-batch size of one, the total amount of computation in the network is 69.33 Gflop, and the network requires 28.15 MB of parameters. 

The execution of CosmoFlow starts with python and the supporting TensorFlow library being  launched on each compute node.  
Once the neural network is constructed in TensorFlow, the initial model parameters are broadcast from rank 0 to all other ranks.  
This ensures all ranks start with the identical model as required for data parallel training.  Each rank then enters a loop over epochs, where an epoch consists of training and validation loops.  
The number of iterations of each loop is defined by $N_{iters} = N_{samples} / n_{ranks}$, where $N_{samples}$ is the number of either training or validation samples.  
Dedicated I/O threads in each rank buffer randomly selected samples into memory from disk for both training and validation.  
The training loop consists of gradient calculation, gradient averaging via MPI communication, and model update from the globally averaged gradients.  The validation loop consists of loss calculation and global averaging.

\subsection{ Single-node results}\label{sec:single}

We first present our analysis of single-node performance on one KNL node of the Cori system with a single MPI rank and batch size of one. 
As we scale, communication starts taking more and more time, but the compute fraction of the profile on a node at scale should stay similar to the single-node one because we keep mini-batch size per node fix.
We enable the CPE ML plugin even at the single node and run with real data from the burst buffer.
We used the profiler to collect sampling data for 100 iterations, starting at iteration 10 and ending at iteration 109 (thus allowing for warmup). 
The average elapsed time per iteration is 145 ms (profiling adds 15 ms overhead).

The majority of the floating-point operations occur in the forward and backward convolution layers. Table~\ref{convperf} shows the  performance of each convolution layer; 
the larger convolutions achieve well over 1 TFlop/Second. 
Many of the smaller convolutions do not, but their contribution to the total time is also less. 
The convolutions were timed with print statements and this caused noticeable overhead on Cori, so the total time shown in Table~\ref{convperf} is greater than the time shown in Figure~\ref{fig:topoperf}.
\begin{table}
\caption{Convolution Layer Performance\label{convperf}, showing the time and performance of each layer for forward data (Fwd), backward weights (Bww), and backward data (Bwd)}
\begin{tabular}{l | r r r | r r r }
Layer & \multicolumn{3}{c}{Time(ms)} &\multicolumn{3}{c}{TF/Sec} \\
& Fwd & Bww & Bwd & Fwd & Bww & Bwd \\
\hline
Conv3D\_1  &     1.14	&	0.74	&		    &	1.52	&	2.32	&	        \\
Conv3D\_2  &     4.04	&	6.20	&	6.76	&	3.51	&	2.28	&	2.09    \\
Conv3D\_3  &     2.32	&	2.65	&	2.84	&	2.22	&	1.95	&	1.82    \\
Conv3D\_4  &     0.40	&	0.39	&	0.42    &	0.25	&	0.25	&	0.22    \\
Conv3D\_5  &     0.32	&	0.29	&	0.40	&	0.35	&	0.37	&	0.31    \\
Conv3D\_6  &     0.22	&	0.29	&	0.30	&	0.10	&	0.08	&	0.09    \\
Conv3D\_7  &     0.18	&	0.22	&	0.21	&	0.04	&	0.03	&	0.05    \\
\hline
{\bf Total} &   8.62	&	10.78	&	10.94	&	2.47	&	1.97	&	1.79

\end{tabular}
\end{table}

Figure~\ref{fig:topoperf} shows the time breakdown by computational units of the single node run. The run used 64 OpenMP threads (including the master thread), 6 I/O threads, and 4 CPE ML Plugin threads. The three bars represent compute profiles of master, worker and communication threads respectively.
In this single-node run, only one of the I/O threads was active, and the plugin threads spent most of their time spinning (since there is no actual communication in this case).
Since the node has only 68 hardware cores, some cores were assigned more than one thread. Linux scheduling worked well for this application and scheduled the active I/O thread and the OpenMP master thread on the same core. This is the Master column in the figure. The 63 OpenMP worker threads, shown as Worker, and the 4 plugin threads, shown as Comms, were also given separate cores.

The CosmoFlow network executes many  element-wise and data reordering operations which are not compute intensive and adds very little to the flops count. Source of the element-wise operations are the forward and backward passes of leaky Relu, (which involves calling two Relu and ReluGrad operations), the Adam+LARC optimizer with polynomial decay, and loss calculations. Each of these functions are threaded separately using loop-level parallelism with OpenMP. 
A load imbalance is created among OpenMP threads due to the low compute load in these functions, and the speedup from OpenMP is less than the ideal case. Also, data reordering between the blocked and non-blocked layout occur at various stages of the graph execution. 

We achieve 535 Gflop/s performance on a single KNL node including the overhead of I/O and the CPE ML Plugin. 
We also note that the corresponding performance on a single GPU node f Piz Daint system is 388 Gflop/s.

\begin{figure}
\includegraphics[width=\columnwidth]{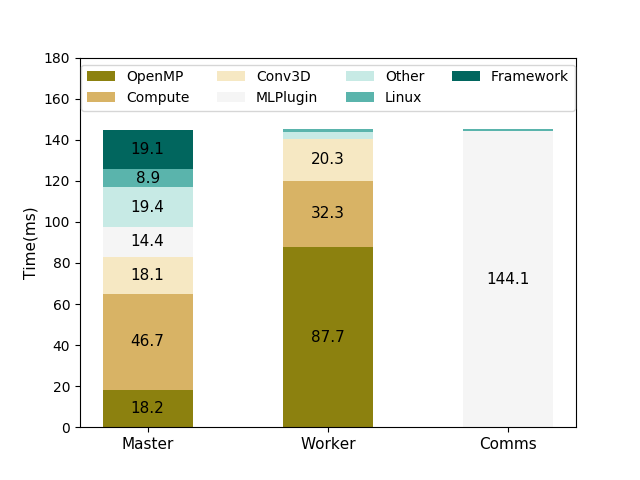}
\caption{Profile of time spent in various stages of the CosmoFlow application on a single KNL node. The stages are OpenMP spin time and overhead, non-convolutional computational time, 3D convolutions, CPE ML Plugin, other time, Linux kernel time, and TensorFlow framework time.}
\label{fig:topoperf}
\end{figure}

\subsection{Multi-node scaling results}\label{sec:scaling}

We perform scaling runs on Cori and Piz-Daint systems described in figure~\ref{fig:coriPizScale}. As we scale on multiple nodes, batch size per node is kept constant at one. Therefore, global batch size is equal to the number of nodes.  
The performance of the multi-node scaling runs is measured in terms of throughput (walltime per epoch).  Reported speedups are the walltime speedups relative to the time taken in a single epoch for training on one node.  This includes training loop and loss averaging across MPI ranks at the end of the epoch. The average epoch time is taken from 8 epochs in a 10 epoch run.  The first two epochs are not included in the average. 
These measurements capture the end-to-end capability of the system and software, including the single node optimizations, efficiency of the communication approach, I/O and interconnect subsystems.

\subsubsection{Cori}
The first set of throughput measurements on Cori are with the training data staged on the Lustre filesystem.  The right plot of Figure \ref{fig:coriPizScale} shows poor scaling beyond 512 nodes with efficiency dropping to less than 58\% at 1024 nodes.  
To investigate this scaling drop, we  perform tests with dummy data (i.e. data not read from a filesystem and instead generated during compute) which suggest that I/O causes significant scaling drop. 

To overcome the read bandwidth bottleneck, the training data is next placed on the DataWarp ``burst buffer" (BB) filesystem on Cori. With higher read bandwidth from the BB filesystem, we see improved scaling efficiency for all the node count as shown in the left part of figure~\ref{fig:coriPizScale}. We achieve scaling efficiency of 77\% on 8192 KNL nodes as shown in the figure.

\subsubsection{Piz Daint}
The throughput measurements are repeated on Piz Daint.  The right plot of Figure \ref{fig:coriPizScale} shows the results with training samples placed on the Piz Daint Lustre filesystem.  Just as was observed on Cori, a probable read bottleneck is encountered at 512 nodes and beyond. The scaling efficiency drops to 44\% at 512 node count. This is in-line with the scaling drop seen on Cori with the data on Lustre filesystem. 
Similar to Cori, the Lustre filesystem is a heavily used shared resource on Piz Daint.  
The DataWarp filesystem on Piz Daint is expected to give better scaling, which we will investigate in future work.

\subsection{Full scale run}\label{sec:full}
Our biggest run uses 8192 KNL nodes of Cori, completing a total of 130 training epochs. 
At this scale, every process sees 20 samples per training epoch. 
The Cori system remains highly stable throughout the run, with the an average epoch time of 3.35 seconds with a standard deviation of $\pm0.32$ seconds (not counting the first epoch). 
The entire run took roughly 9 minutes total with 8 minutes of training time. 
We achieve an average sustained performance of slightly over 3.5 Pflop/s single precision for 8192 nodes with a parallel efficiency of 77\% 
relative to a single node (6324X speedup).  This speedup is based on the timing of a full epoch.  We achieved 80\% scaling efficiency using only the step time (excludes validation and loop overheads).

We also perform a convergence run on 2048 nodes, with a sub-set of the dataset used to train the 8192-node attempt. 
Our efficient and scalable software stack allows us to tune hyper-parameters such as base learning rate, minimum learning rate and decay epochs for both of these runs and experiment with variations in datasets. The loss function for both the 2048-node and 8192-node runs are shown in figure~\ref{fig:loss8192}. 
The network clearly converges with fewer number of epochs in the 2048-node run. We discuss the implications of this result for convergence at scale in Section~\ref{sec:convergence}. 

\begin{figure*}[ht]  
\centering
\begin{minipage}{0.99\columnwidth}
  \includegraphics[width=.9\columnwidth]{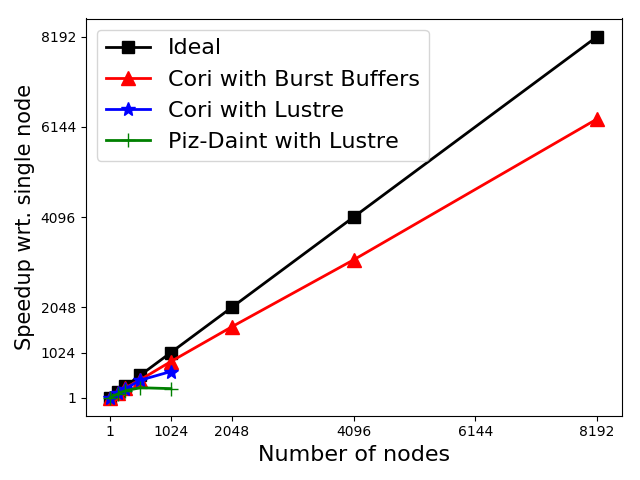}
\end{minipage}%
\begin{minipage}{0.99\columnwidth}
  \includegraphics[width=.9\columnwidth]{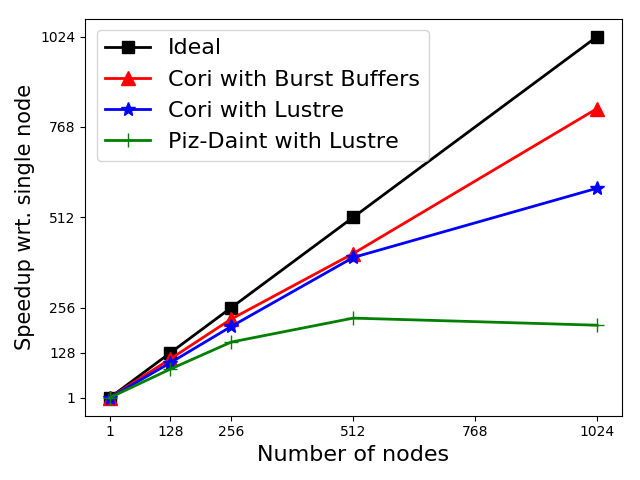}
\end{minipage}
  \caption{Scaling of fully synchronous training on Cori and Piz-Daint based on epoch timing. Right figure is a zoomed-in version of the left one, highlighting the scaling efficiency drop from Lustre file system.}
\label{fig:coriPizScale}	
\end{figure*}

\section{Discussion }
\label{sec:discussion}

\subsection{I/O}

Comparing the scaling performance between training data staged on Lustre and then on DataWarp, we can examine the read performance needed to obtain high efficiency.  The CosmoFlow code uses the $QueueRunner$ and $coordinator$ features of TensorFlow to read and buffer training samples in a pipeline behind gradient computation.  Ideally this should hide the cost of I/O as long as there is sufficient read bandwidth.  We can estimate the minimum read bandwidth necessary to hide I/O costs from
\begin{align}
BW_{min} (MB/s/node) = \frac{b \times S(MB)}{t(s)},
\label{eqn:bw}
\end{align}
where $b$ is the mini-batch size per process, $S$ is the sample size in MB, and $t$ is the step time at that batch size. With $b=1$, $S=8$ MB and $t\approx$ 0.129 seconds, the minimal required read bandwidth per compute node is 62 MB/s.  With a maximum bandwidth of 700 GB/s for 248 OSTs on Cori Lustre filesystem, each OST should be capable of 2.8 GB/s and be able to feed 46 compute nodes fast enough for CosmoFlow.  We find that absolute performance is 16\% better using DataWarp at 128 MPI ranks, which suggests I/O is already a bottleneck at that scale.

The step time at 128 nodes is 150 ms using DataWarp and 179 ms using Lustre.  Assuming I/O from Lustre limits performance, Equation \ref{eqn:bw} predicts an average of 90 MB/s from each of the 64 OSTs.  There are several reasons to expect less than peak bandwidth from each OST including read location on each spinning disk, diversity of OSTs used in a moment across all nodes, and that it is a shared resource on the system.  We suspect there is a wide range in bandwidth actually being delivered across the OSTs and that the measured performance is limited by the lowest bandwidth or significant contention.

\subsection{Communication}

We can estimate the latency of the gradient aggregation communication with the CPE ML Plugin by comparing the step time at 1024 nodes to 1 node on Cori.  A single node with training samples read from DataWarp achieves 7.72 samples/sec or a step time of 129 ms.  At 1024 nodes, each node achieves 6.19 samples/sec or a step time of 162 ms.  The latency from gradient aggregation is 33 ms assuming that it alone accounts for the difference in step times.  Noting that the reduction algorithm communicates twice the message length for large MPI rank counts, an estimate for the achieved bandwidth from communication is ($2 \times 28.15$ MB) / 0.033 sec = 1.7 GB/s/node.  Each node for the 8192 node job achieved 5.96 samples/sec or a step time of 168 ms. 
The estimated performance of the CPE ML Plugin at 8192 nodes is therefore 1.42 GB/s/node.

The Aries interconnect is capable of $\approx$ 10 GB/s/node unidirectional between any two endpoints.  These results show the CPE ML Plugin can achieve a high bandwidth even with math operations interleaved and global communication.  It also shows the effectiveness of the CPE ML Plugin at hiding any ``straggler'' effects.



\section{Implications}
\label{sec:implications}

\subsection{Cosmology results}
\label{sec:convergence}
\begin{figure}
\begin{minipage}{0.5\columnwidth}
  \includegraphics[width=.99\columnwidth]{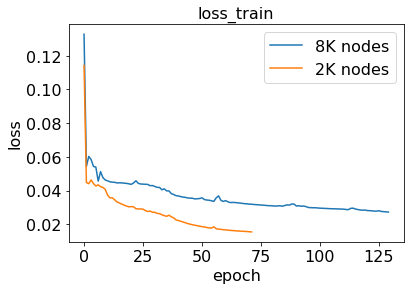}
\end{minipage}%
\begin{minipage}{.5\columnwidth}
  \centering
  \includegraphics[width=.99\columnwidth]{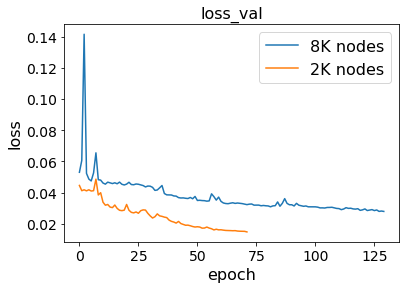}
\end{minipage}
  \caption{Loss function for the training dataset (left) and validation dataset (right), for 2048 and 8192 node runs. Note that the datasets used in the two runs are not same and of different size, so the loss function and epoch time cannot be directly compared.    }
\label{fig:loss8192}	
\end{figure}

Figure~\ref{fig:Sigma8-OmegaM-Ns} shows the parameter estimates from the 2048- and 8192-node runs for the CosmoFlow network predicting three cosmological parameters.
We calculate the average relative error of the parameter estimation using $( |\Omega_{M, model} - \Omega_{M, true}| ) / \Omega_{M, model}$ (where $\Omega_{M, model}$ is the model estimate and $\Omega_{M, true}$ is the true value). We obtain relative errors of (0.0022, 0.0094, 0.0096) for ($\Omega_M, \sigma_8$, $n_s$) with the 2048-node run. 
This is comparable to the best experimental uncertainty~\cite{Planck2016,DESY1} for $\Omega_M$ and $\sigma_8$, and almost 5$\times$ smaller for $n_s$. 
These relative errors are significantly lower than those obtained by Ravanbakhsh et al. (2017)~\cite{siamak2017} using their CNN where they only predicted two parameters. We attribute this improvement in accuracy to the larger simulation volume, despite the network being asked to infer an additional parameter. 
We note that experimental observations using the dark matter distribution are generally not able to measure $n_s$ accurately; this work demonstrates that such a measurement can be made possible using advanced machine learning techniques. 

It is clear that the 8192-node run, although learning, is not converged to the same accuracy as the 2048 node target run.  
However we still see good parameter estimation, obtaining a relative uncertainty of (0.052, 0.014, 0.022)  for ($\Omega_M, \sigma_8$, $n_s$), within a factor of 2 of today's best experimental results~\cite{Planck2016,DESY1}. We believe that with additional hyper-parameter and optimizer tuning at 8192 nodes, it is possible to achieve similar accuracy levels as the 2048 node target run.

\begin{figure}
\begin{minipage}{.33\columnwidth}
  \centering
  \includegraphics[width=.99\columnwidth]{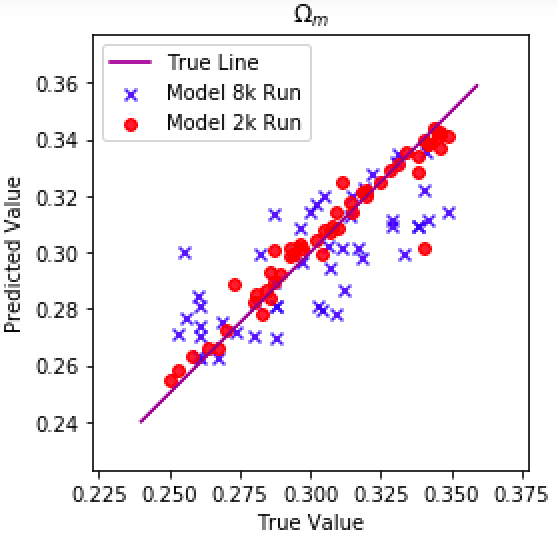}
\end{minipage}%
\begin{minipage}{.33\columnwidth}
  \centering
  \includegraphics[width=.99\columnwidth]{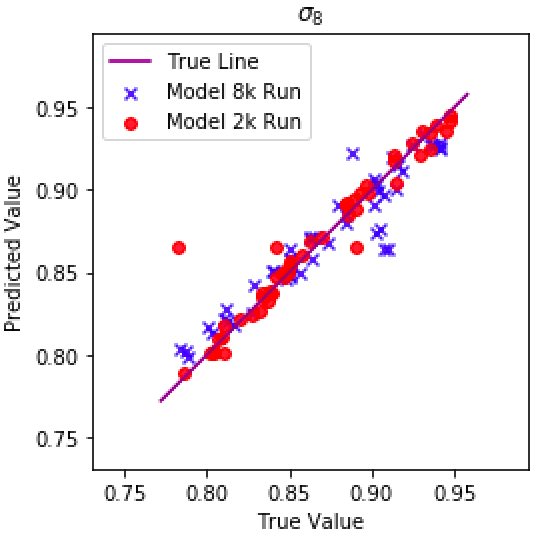}
\end{minipage}%
\begin{minipage}{.33\columnwidth}
  \centering
  \includegraphics[width=.99\columnwidth]{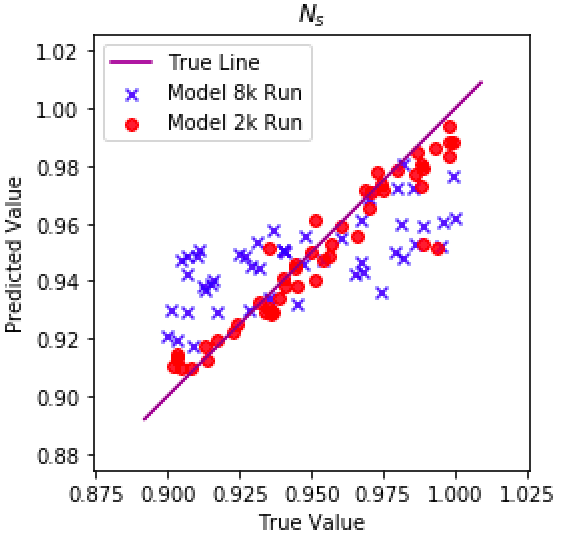}
\end{minipage}
\caption{Estimates of $\Omega_M$, $\sigma_8$ and $n_s$ from the 2048- and 8192-node runs.}
 \label{fig:Sigma8-OmegaM-Ns}	
\end{figure}

\subsection{Implications on the application and future hardware }

With the current CosmoFlow framework, we achieve reasonably accurate predictions on 8192 nodes in just a few minutes. 
This opens up new avenues for exploration of extended cosmological problems. 
For example, extending the network to predict more cosmological parameters, designing optimized hyperparameter searches, extending the network to multiple redshift snapshots or combining multiple experimental probes are now within the reach. 
CosmoFlow enables fast, efficient exploration of the possibilities deep learning can offer cosmology, and indeed other science areas with 3D data.

With the large amount of data being processed for deep learning applications, efficient I/O hardware and software becomes critical piece for obtaining performance. Our scaling experiments clearly demonstrated a significant efficiency drop on the Lustre 
file system, which we are able to compensate by using a high bandwidth SSD-based file system. 

Besides the I/O challenge, deep learning also poses a supercomputing challenge at Exascale. 
The three-parameter estimation problem using $128^3$ voxel dataset size would require more than 60 days of execution time on a single node to converge to a model at an acceptable loss.  
Training times of this scale are impractical, which is why developing efficient HPC-based techniques for scaling deep learning is critical.  
There are limitations to the methods described in this work, however.  Data parallel training methods, such as SSGD, are fundamentally limited by the size of the training dataset.  
The dataset must have substantially more samples in it than the target level of concurrency (i.e. MPI ranks).
Furthermore, multiple runs are required to tune the hyper-parameters in the problem.  
With these challenges, the only way to continue pursuing scientific discoveries with deep learning is to develop highly efficient single node and scalable software for supercomputing resources. 
Applying HPC optimization techniques, we are able to optimize single-node performance at high compute efficiency and scale the training problem to 8192 nodes on Cori with 77\% scaling efficiency.

\section{Conclusions}
\label{sec:conclusion}
In this paper, we presented an optimized and highly scalable deep learning application CosmoFlow, which is built on top of the TensorFlow framework and predicts 3 cosmological parameters. 
We optimize the full software stack which includes network design, I/O processing pipeline, communication, TensorFlow framework and 3D CNN primitives in MKL-DNN for our 3D convolutional neural network on the Cori supercomputer. 
We discuss the importance of I/O hardware and software stack for deep learning applications with the scaling runs on both the Cori and Piz-Daint systems. 
We also highlight challenges in achieving convergence at scale with a fully synchronous SGD algorithm.  
We processed over 1.4TB of data on 8192 nodes, reaching 535 Gflop/s on a single KNL node and sustaining 3.5 Pflop/s in aggregate on Cori. 
The configuration shows good scaling with 77\%\ parallel efficiency on up to 8192 nodes. 
We achieve full convergence at 2048 nodes with best in-class scientific results. We also achieve close--to--converged results on 8192 nodes, which produce reasonable scientific predictions.

The success of deep learning for scientific problems will hinge upon our ability to develop and optimize algorithms and implementations that can handle the complexity of scientific data efficiently. 
In this work we address the issue of multi-dimensional data, using 3D simulations of   matter distribution to solve a pressing problem in cosmology - how to determine the parameters that describe the nature of the universe. 
Our scaling and performance improvements have enabled us to achieve an unprecedented level of accuracy in our estimation of cosmological parameters - in particular, we are able to accurately estimate $n_s$ using only the dark matter distribution.

\section*{Disclaimers}
\footnotesize
\noindent Software and workloads used in performance tests may have been optimized for performance only on Intel microprocessors. 

\noindent Performance tests, such as SYSmark and MobileMark, are measured using specific computer systems, components, software, operations and functions. Any change to any of those factors may cause the results to vary. You should consult other information and performance tests to assist you in fully evaluating your contemplated purchases, including the performance of that product when combined with other products. For more complete information visit www.intel.com/benchmarks. 

\noindent Performance results are based on testing as of April 13, 2018 and may not reflect all publicly available security updates. See configuration disclosure for details. No product can be absolutely secure. 

\noindent Configurations: Testing on Cori (see \S\ref{sec:cori}) was performed by NERSC, with the spectre\_v1 and meltdown patches. Testing on Piz Daint (see \S\ref{sec:piz-daint}) was performed by Cray.

\noindent Intel does not control or audit third-party benchmark data or the web sites referenced in this document. You should visit the referenced web site and confirm whether referenced data are accurate. 

\noindent Intel, VTune, Xeon, and Intel Xeon Phi are trademarks of Intel Corporation or its subsidiaries in the U.S. and/or other countries.

\bibliographystyle{IEEEtran}
\bibliography{cosmoflow}

\end{document}